\documentclass[aps,pre,twocolumn,floatfix,amsmath,showpacs]{revtex4}\usepackage{ae}
\usepackage{graphicx}
\usepackage{bm}
\usepackage{wasysym}
\usepackage{epsf}
\usepackage{epsfig}
\usepackage{amssymb}
\usepackage{color}
\DeclareGraphicsExtensions{.pdf,.png,.jpg} 
\begin{document}
\newcommand{\wq}[1]{\textcolor{blue}{#1}}

\title{
Log-periodic oscillations for diffusion on self-similar finitely ramified structures 
}
\author{L.~Padilla}
\thanks{Fellow of Consejo Nacional de Investigaciones Cient\'{\i}ficas y T\'ecnicas (CONICET)}
\author{H.~O.~M\'artin}
\thanks{Research Member of CONICET}
\author{J.~L.~Iguain$^\dag$}
\email{iguain@mdp.edu.ar}
\affiliation{Instituto de Investigaciones F\'{\i}sicas de Mar del Plata (IFIMAR) and
Departamento de F\'{\i}sica FCEyN,\\
Universidad Nacional de Mar del Plata, De\'an Funes 3350, 7600 Mar del
Plata, Argentina}


\begin{abstract}
Under certain circumstances, the time behavior of a random walk is 
modulated by logarithmic periodic oscillations. The goal of this paper is
to present a simple and pedagogical explanation of the origin of this
modulation for diffusion on a substrate with two properties: self-similarity and finite ramification order.  
On these media, the time dependence of the mean-square displacement shows 
log-periodic modulations around a leading power law, which can be understood on the base
of a hierarchical set of diffusion constants.
Both the random walk exponent and the period of oscillations are analytically obtained for a pair 
of examples, one fractal, the other non-fractal, and  confirmed by Monte Carlo simulations.   
\end{abstract}
\maketitle
\section{
Introduction
\label{intro}
}

Diffusion on non-Euclidean media is not a new research area. For example, the underlying mechanisms of the 
sub-diffusive behavior, characteristic of fractals, were discovered some decades ago. Indeed, it is by now
well established that, on these objects, the spreading of the probability density function is continuously retarded because
of the presence of holes of all sizes (the interested reader can refer
to~\cite{ale,ram,ben,hav,bou},  and references therein).
However, in the last years, considerable effort has been dedicated to investigate an outstanding phenomenon. It has been repeatedly  
reported that, on some deterministic fractals or self-similar graphs, the time behavior of a random walk (RW) is modulated 
by logarithmic-periodic oscillations~\cite{gra,kro,ace,bab1,bab2,maltz}. Similar effects were observed for biased diffusion
of tracers on random systems \cite{bernas,stau0,stau,yu}, and out of the domain of diffusive motion, examples have been detected in 
earthquakes \cite{huang,saleur}, escape probabilities in chaotic maps close to crisis \cite{pola},
kinetic and dynamic processes on random quenched and fractal 
media \cite{kut,andra,bab3,saleur2}, diffusion-limited 
aggregates \cite{sor1i}, growth models \cite{huang2}, and stock markets near a 
financial crash \cite{sor2,van1,van2,van3}.
There is  general agreement that these oscillations are a manifestation of an inherent 
self-similarity \cite{dou}, responsible for a discrete scale invariance \cite{sor1}.

A wide class of systems exhibiting log-periodicity is that of 
{\it self-similar finitely ramified} structures.
Indeed, several researchers have found that, on these substrates, the  root-mean-square displacement (RMSD) 
of a single RW  as a function of time, follows an anomalous power law modulated by logarithmic periodic oscillations
\cite{bab1,bab2,maltz} (for related behavior, see \cite{gra,kro,ace} and \cite{dou}).
In this paper, we revisit this problem, with the purpose of explaining  the origin of the behavior described above in a simple and
comprehensible way. 
The work is a natural continuation of a previous one \cite{lore1}, where we studied log-periodic modulation in
one-dimensional RW.

Our approach is two-fold: theoretical analysis and numerical simulations.
In the theoretical part, we treat the diffusion problem in a quite general way. 
The analysis is facilitated by the fact that a part of a finitely ramified object can be separated from 
the rest by cutting a finite number of connections, which, because of self-similarity, does not depend on the size of 
the part to remove. 
For pedagogical reasons, the  procedure is illustrated by two simple examples but the                                           
method should be easily generalized for other finitely ramified substrates.
 The first model of substrate, denoted I, is the well-known Vicsek's {\it snowflake} fractal structure~\cite{vic}.
The second, Model II, is also finitely ramified but may be  considered a {\it trivial} fractal 
(with a fractal dimension of $2$ and without holes of all sizes). Model I is an example of what we will call a {\it perfect diffusive
self-similar} structure; Model II corresponds to what we will call  an {\it asymptotically diffusive self-similar} one.
For both models we predict a time-modulated oscillatory behavior, and calculate the RW exponent and the period of the oscillations.
The goal of the numerical part is to get some independent confirmation of the derived properties.
We study, using standard Monte Carlo (MC) simulations, the time evolution of a particle diffusing on each substrate, and compare the findings
with our theoretical predictions.

The paper is organized as follows. In Sec.~\ref{analy} we define the models, discuss the causes for modulated power-law behavior 
and derive the RW time evolution. The outcome of MC simulations are presented in Sec.~\ref{nume}. Finally, we give our conclusions
in Sec.~\ref{conclu}.

\section{Analytic Approach
\label{analy}
}
The above mentioned substrates are built in stages on a square lattice and the 
result of every stage is called a {\it generation}. 
The structure of each generation is a periodic array of basic or unit cells, which consist of sites connected by
bonds. We denote by $L$ the linear size of the unit cell that corresponds to the first generation.

On any generation, the motion of a single particle, initially located at a symmetry point of the structure, occurs stochastically. 
The particle jumps, with a hopping rate $k$, only  between NN substrate sites which are connected by a bond.
 
\subsection{
Model I
}
\begin{figure}
\includegraphics[width=\linewidth]{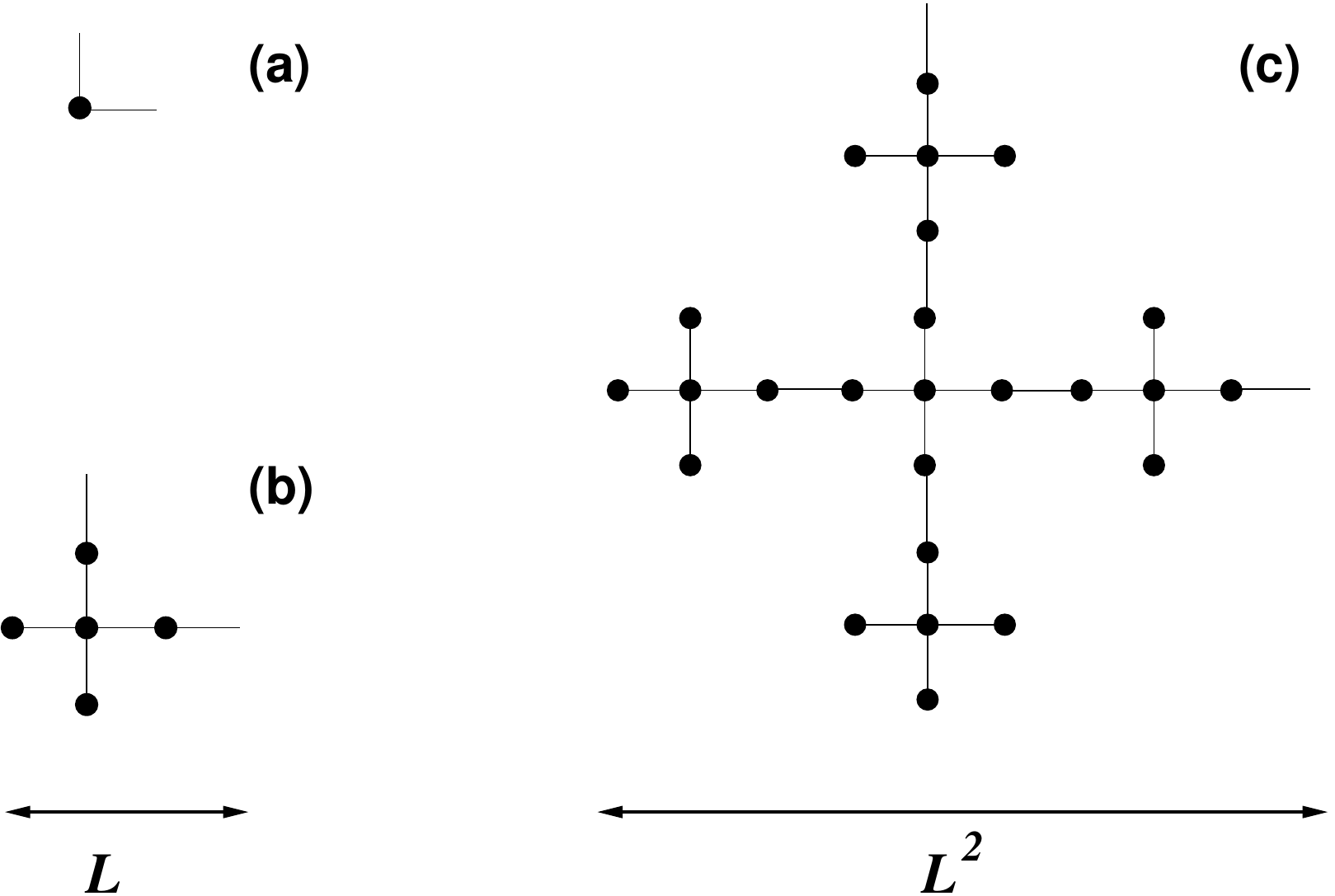}
\caption{The unit cells of Model I. The zeroth, first, and second generations are shown in (a), (b), and (c), respectively. 
$L$ (3 in this case) is the linear size of the first generation. 
}
\label{mo1}
\end{figure}

The building process is illustrated in Fig.~\ref{mo1}. Parts (a), (b) and (c) of this figure show the unit cell for
the zeroth, first and second generation, respectively. It is easy to see that, for this model, $L=3a$, where $a$ is the 
distance between nearest-neighbor (NN) sites ($a=1$ in the rest of the article). 
It is also apparent from this figure that the second-generation basic cell has a linear size $L^2$ and is built from  
the first-generation one in a self-similar way.
In general, the $n$-th generation basic cell has a linear size $L^n$ and, for $n>1$, a central part of a linear size $L^{(n-m)}$ 
(with $m=1,2,...,n-1$) can be separated from the rest  by cutting only four bonds. 

\begin{figure}
\includegraphics[width=0.7\linewidth]{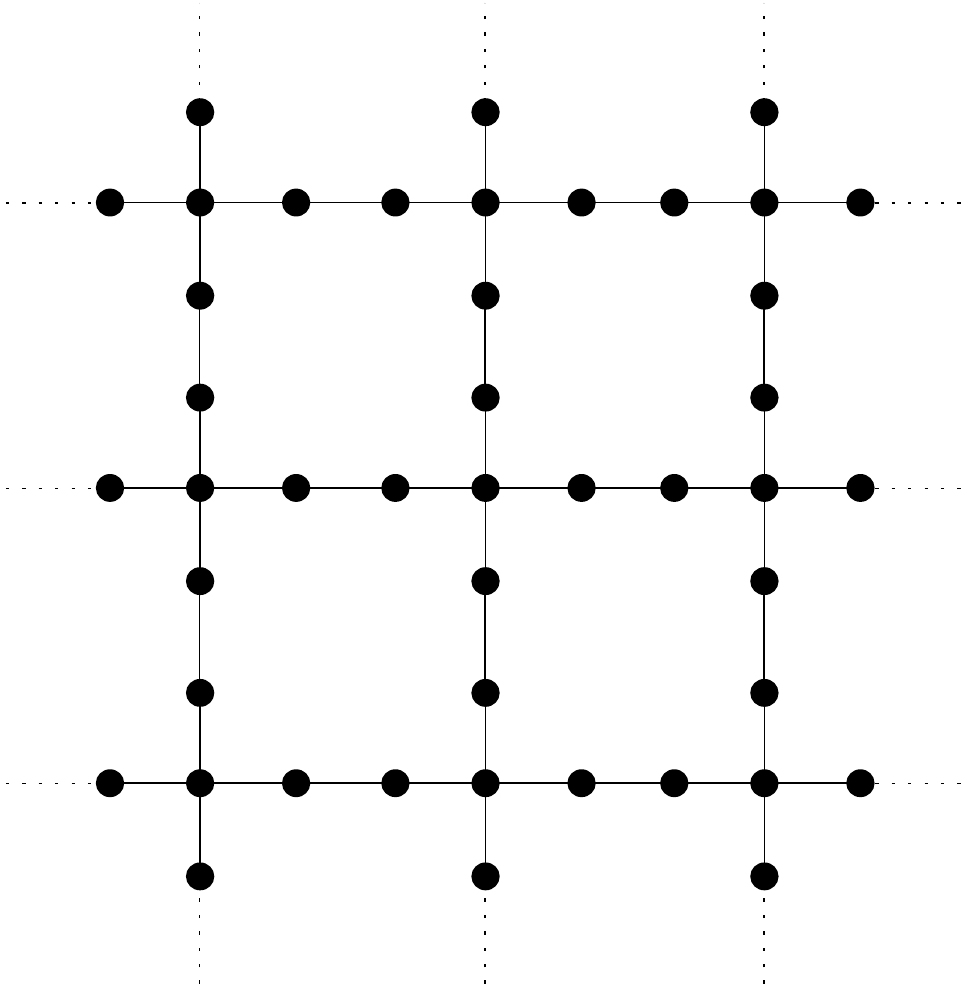}
~\vspace*{1.5cm}\\
\includegraphics[width=\linewidth]{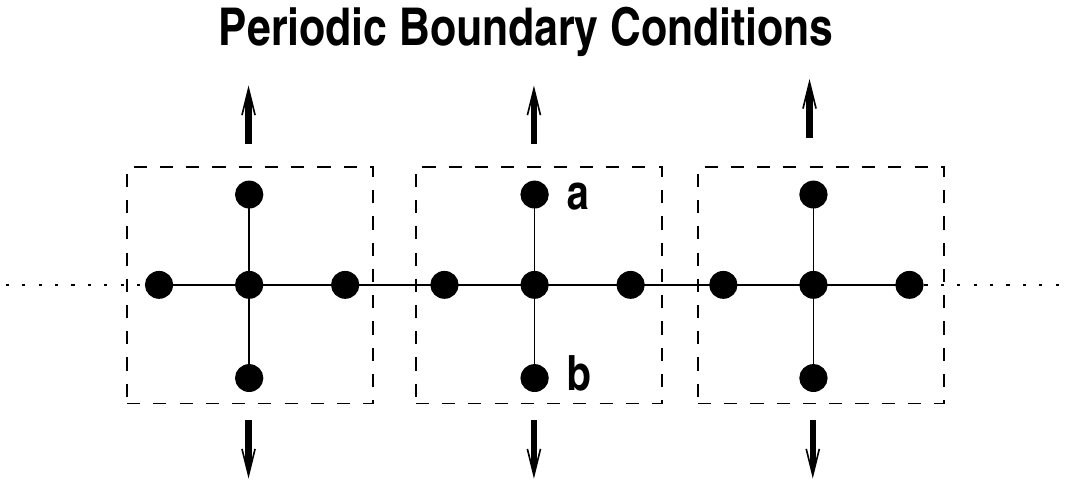}
\caption{
First generation of Model I. Top: the substrate built with the basic cell shown in Fig.~\ref{mo1}-(b).
Bottom: the infinite one-dimensional string of cells used to compute the diffusion coefficient $D^{(1)}$ (see Fig.2
in Ref.~\cite{celso}).  The arrows indicate 
periodic boundary condition in the $y$ direction meaning that when, for example, the RW at site {\bf a} ({\bf b}) jumps 
upward (downward), it will arrive at site {\bf b} ({\bf a}).
}
\label{strip}
\end{figure}

For the $n$-th generation, a two-dimensional periodic substrate is built by connecting the corresponding basic cell
(the first-generation substrate can be observed at the top of Fig.~\ref{strip}). 
The final full self-similar substrate, we are interested in, is obtained after 
an infinite number of iterations. Note that, in this limit, the unit cell is identical to the snowflake fractal.

We proceed now to analyze the behavior of the diffusing particle. It is useful to remember that, on any periodic substrate,
normal diffusion should be observed if time is long enough for the RW to be influenced by the structure periodicity.
Thus, for the $n$-th generation substrate, a diffusion coefficient $D^{(n)}$ can be defined 
through the time dependence of the mean-square displacement  \mbox{$\langle\Delta^2 x\rangle(t) = \langle[x(t)-x(0)]^2\rangle$}
in the $x$ direction,

\begin{equation}
\langle\Delta^2 x\rangle(t) = 2 D^{(n)} t,
\label{evo}
\end{equation}
which holds for a time $t$ longer that the average time for the particle to escape from the initial
unit cell. Because of the $x\leftrightarrow y$ symmetry, the same time dependence holds for the
mean-square displacement in $y$ direction. 

It is not hard to convince oneself that the  zeroth-generation substrate is the
simple square lattice (see  Fig.~\ref{mo1}-(a)), and then

\begin{equation}
              D^{(0)}= k.
\label{homo}
\end{equation}

The first-generation substrate is shown at the top of Fig.~\ref{strip}. However, 
regarding $x$-direction diffusion, the whole substrate
and the string of cells displayed at the bottom of Fig.~\ref{strip}, with periodic boundary conditions in
the $y$-direction, lead to equivalent problems. We exploit this equivalence and calculate the diffusion coefficient of that
one-dimensional array of equivalent cells. Following the steady-state method of Ref.~\cite{celso}, we obtain

\begin{equation}
 D^{(1)}= (3/5) k. 
\label{first}
\end{equation}

The reduction to a quasi one-dimensional problem is 
possible for every $n$, and, after some simple algebra, it is found that   

\begin{equation}
       D^{(n)}= (3/5)^n k, \;\;\;\mbox{for}\; n=0, 1, 2,....
\end{equation}

On the base of this result, we can anticipate that, on the full self-similar structure, a RW
will show a log-periodic modulated behavior. The key point here is that the diffusion coefficients satisfy 

\begin{equation}
D^{(n)}/D^{(n+1)}=\delta,  \;\;\;\mbox{for}\; n=0, 1, 2,...,
\label{quotient}
\end{equation}
with  $\delta= 5/3 $.

Let us start noting that, since at a time $t$ the particle will be located in a region of linear size of the order of 
the RMSD $\sqrt{\langle\Delta^2 x\rangle(t)}$, for a time such that 
 $A L^n < \sqrt{\langle \Delta^2 x\rangle(t)} < A L^{n+1}$ ($A$ is a constant of the order of one), it will be impossible for 
the RW to distinguish the full self-similar substrate from the
$(n+1)$-th-generation one. Everything will thus happen as on the latter and one should expect that 

\begin{equation}
\langle\Delta^2 x\rangle(t) =  2 D^{(n)} t,
\end{equation}
for $t$ in that interval.

At later times, when the RMSD is of the order of  $L^{n+1}$, 
the particle will start to diffuse as if located 
on the $(n+1)$-th generation substrate. The relation 

\begin{equation}
  \langle\Delta^2 x\rangle(t) = 2 D^{(n+1)} t
\end{equation} 
will thus hold for $A L^{n+1} < \sqrt{\langle \Delta^2 x\rangle(t)} < A L^{n+2}$.

\begin{figure}[h]
\includegraphics[width=\linewidth]{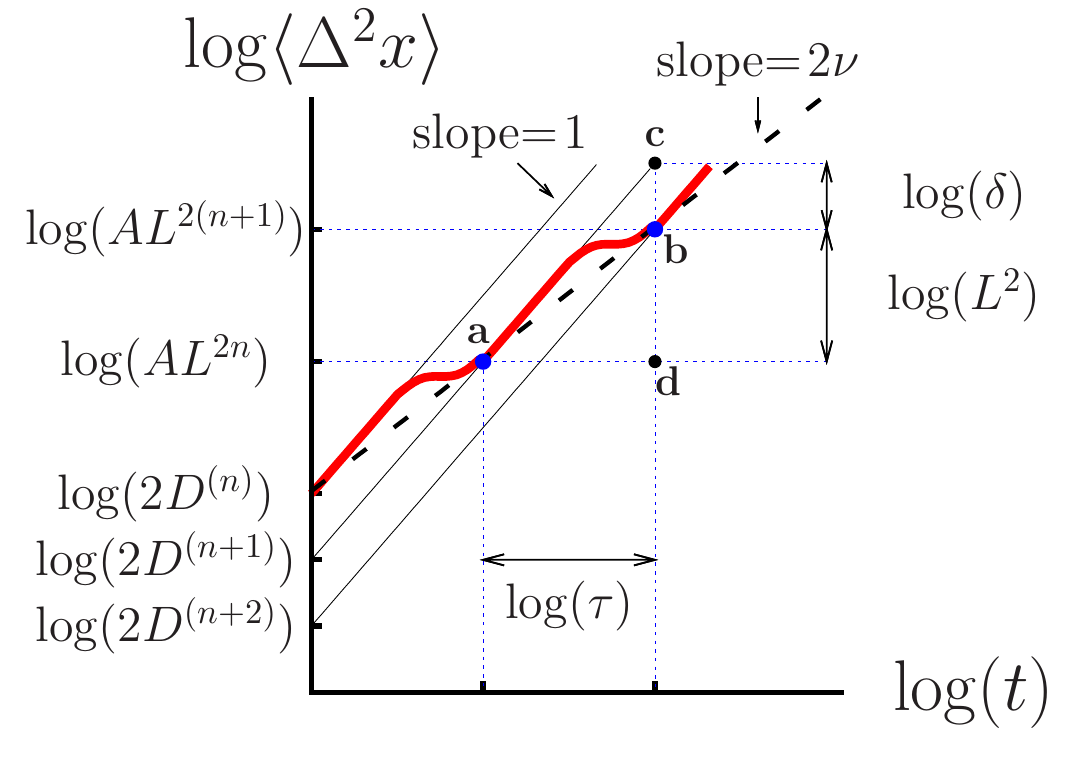}
\caption{(Color online) Schematic of the mean-square displacement
 as a function of the time, shown by the thick red curve. The length
of the segment {\bf bc} is $\log(2D^{(n)})-\log(2D^{(n+1)})=\log(\delta)$,
because of Eq.~(\ref{quotient}). From the slopes ($=1$) of the full straight
lines (representing the normal diffusion behaviors, 
$\langle \Delta^2x \rangle = 2 D^{(n)}t$), one gets that the segments {\bf ad} and
{\bf cd} have the same length or, equivalently,  that
$\log(\tau)=\log(L^2)+\log(\delta)$. The dashed straight line represents the
global power law   $ \langle \Delta^2x \rangle  \sim t^{2\nu}$, with
$2\nu= \log(L^2)/\log(\tau)$. More details in the text.
}
\label{qualita}
\end{figure}

Because of the substrate full self-similarity, we should observe a sequence of 
length-dependent diffusion coefficients, giving rise
to the qualitative behavior depicted in  Fig.~\ref{qualita}. 
This is a sketch of the mean-square displacement as a function
of time (thick curve), which has a power law form  modulated by a log-periodic amplitude. That is,

\begin{equation}
\begin{centering}
\langle \Delta^2x \rangle(t) = C t^{2 \nu} f(t) {\rm ,}
\end{centering} \label{function}
\end{equation}
where $\nu$ is the RW exponent, $f(t)$ is a
log-periodic function which satisfies $f(t \tau)=f(t)$, with
the logarithmic period $\log(\tau)$, and the constant $C$ is obtained 
by asking that the log-time average of $\log(f)$
over one period be zero (see, for more details,
Fig.~\ref{cruz}).

In Fig.~\ref{qualita}, two groups of inclined straight lines were drawn as a guide.
The
dashed line has a slope of $2\nu$ and corresponds to the $\langle  \Delta^2 x\rangle$ global power-law
trend. The solid lines have slopes
of $1$ and represent normal diffusion in each of the different generation structures. 
Both $\tau$ and $\nu$ can be expressed in terms 
of $L$ and $\delta$. It is clear from
 the slopes of the straight lines that
$\log(\tau)=\log(L^2)+\log(\delta)$ (solid line) and
$2\nu=\log(L^2)/\log(\tau)$ (dashed line), which is equivalent to
\begin{equation}
\tau=\delta L^2=L^{1/\nu}\;{\rm ,}
\label{tau}
\end{equation}
and
\begin{equation}
\nu = \frac{1}{2+\frac{\log \delta}{\log L}}\;\;\;  .
\label{exponent}
\end{equation}
Note that, since $\delta>1$, anomalous sub-diffusion results ($\nu<1/2$, see Eq.\ref{exponent}), 
and that, according to the sketch in
Fig.~\ref{qualita}, the amplitude of the modulation
increases with the increase of $\delta$.

We would like to stress again that the self-similarity in the mean-square displacement,
schematically shown in Fig.~\ref{qualita} and mathematically described by
Eq.(\ref{function}), is a direct consequence of (\ref{quotient}). 
It is because of the latter that, in Fig.~\ref{qualita},  the distance between any pair
of nearest solid straight lines is a constant, and that two nearest
equivalent points (like {\bf a} and {\bf b}) are always  related by the transformation
($t\rightarrow\tau t$ ,
$\langle  \Delta^2 x\rangle\rightarrow\tau^{2v}\langle  \Delta^2 x\rangle$). The set of equations (\ref{quotient})
also allows us to obtain analytically both the random walk exponent and the
period of the oscillations. 
We will refer to Model I as a {\it perfect diffusive self-similar} structure.
Other examples of this kind of structures 
are the fractals shown in Fig.(1) of Ref.~\cite{maltz}.

\subsection{
Model II
}

\begin{figure}
\includegraphics[width=\linewidth]{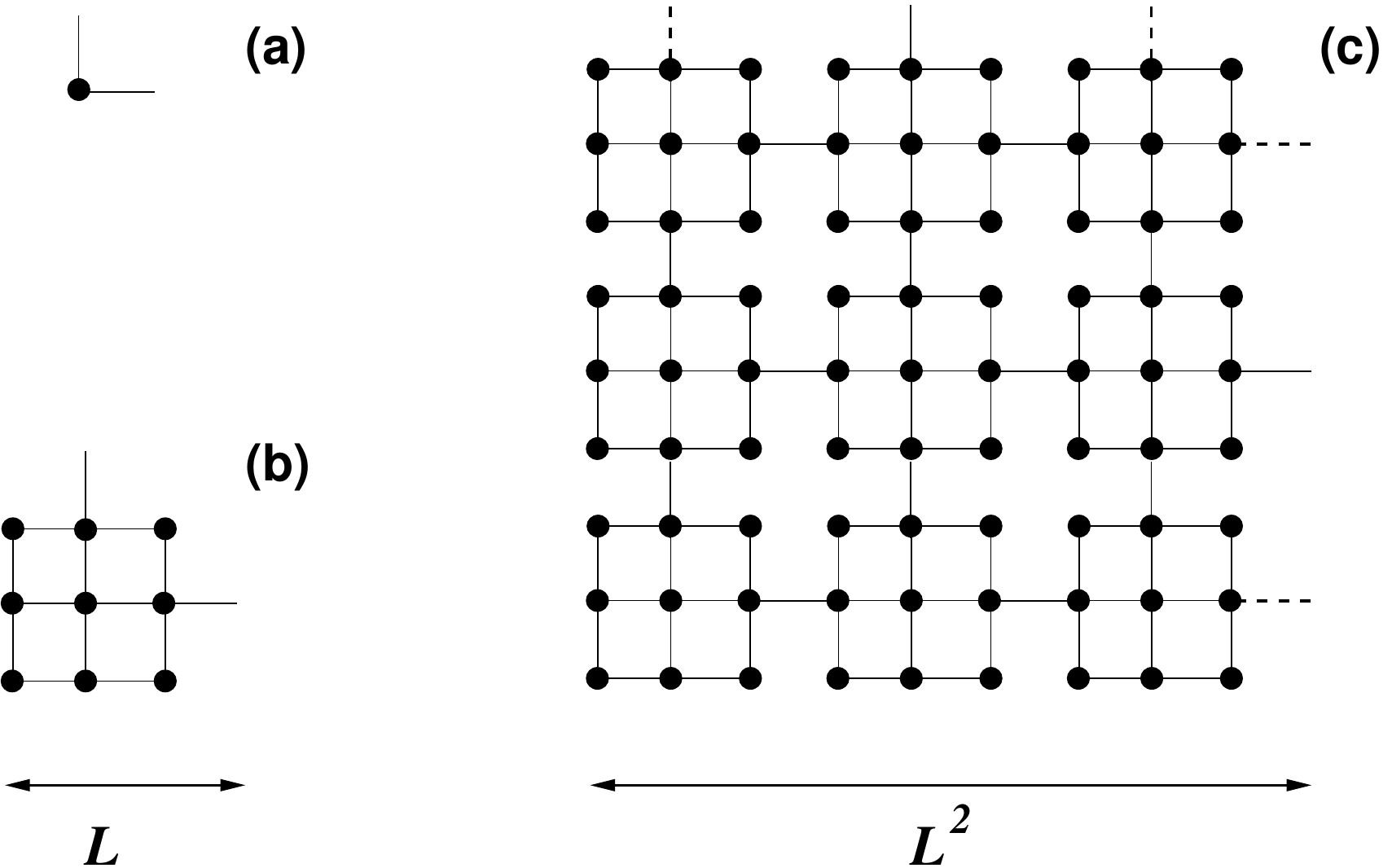}
\caption{The unit cells of Model II. The zeroth, first, and second generations are shown in (a), (b), and (c), respectively.
$L=3$, as in Model I. The dashed lines correspond to the bonds which are removed from the immediately previous generation.
Note that, for every order,  two NN unit cells are connected by only one bond.
}
\label{mo2}
\end{figure}

For this model, the basic cells for the zeroth, first and second generation are shown in Fig.~\ref{mo2}. The 
full self-similar substrate is obtained when the generation order $n$ goes to infinity.
As for Model I, the width of the $n$-th-generation unit cell is $L^n$ (where $L=3$ is the width of the first-generation one), and, 
for $n>1$, a central part of linear size $L^{n-m}$ (for $m<n$) can be separated from the rest by cutting only four bonds.
Again, this allows us to reformulate the RW problem on a one-dimensional array (similar to that shown in Fig.~\ref{strip}-bottom),
and to compute the diffusion coefficients following the steady-state method~\cite{celso}.

It is evident that $D^{(0)}$ is  given by Eq.~(\ref{homo}). For the next two generations we obtain
\begin{equation}
D^{(1)}= (1/2) k,
\label{one}
\end{equation} 
and
\begin{equation}
D^{(2)}= (45/174)  k.
\label{two}
\end{equation} 
We stop here because the calculations become more tedious for higher orders. However, at this point, we are already able to
grasp the general trend, and to make  a comparison between the two models. 

From the values of the first few diffusion coefficients (Eqs.~(\ref{homo}), (\ref{one}) and (\ref{two})), 
we immediately observe that the ratio $D^{(0)}/D^{(1)}= 2$ is different than $D^{(1)}/D^{(2)}= 87/45 \simeq 1.933$.
In general, an analysis of the structures indicates that, for Model II, the equations (\ref{quotient}) should
be replaced by
\begin{equation}
D^{(n)}/D^{(n+1)}=\delta_n, 
\label{quotient_n}
\end{equation}
where, instead of a constant $\delta$, we have now a sequence $\delta_n$. In spite of this modification, the qualitative behavior 
shown in Fig.~\ref{qualita} still holds, but the exponent $\nu$ should be replaced by a sequence 
\begin{equation}
\nu_n = \frac{1}{2+\frac{\log \delta_n}{\log L}}\;\;\;  ,
\label{exponent_n}
\end{equation}
and  the constant period $\tau$ by 
\begin{equation}
\tau_n=L^{1/\nu_n}\;{\rm .}
\label{tau_n}
\end{equation}
It is interesting to note that $\nu_n$ and $\tau_n$ can naturally be interpreted as a local 
RW exponent and a local period, respectively.

Substituting the parameters above ($\delta_0=2$ and $\delta_1=87/45$) into Eqs.~(\ref{exponent_n}) and 
(\ref{tau_n}), one obtains $\nu_0\simeq 0.3801$, $\nu_1\simeq 0.3846$, $\log(\tau_0)\simeq 1.2553$ and $\log(\tau_1)\simeq 1.2406$. 
Even if we do not  calculate $\delta_n$ for every $n$, in the next paragraph
we argue that an oscillating modulation is always present, and that a log-periodic behavior
appears in the long-time limit.
 
Since the above results seem to indicate that $\delta_n$ is a decreasing sequence, 
a crucial question is whether, for large enough $n$, it approaches the trivial limit $\delta_\infty=1$, equivalent
to normal diffusion (see (Eq.~\ref{exponent_n})).
To see that this is not the case, look at Fig.~\ref{mo2} and observe that whenever the distance in the $x$ direction 
from the center of symmetry
exceeds a threshold $L^n/2$, the number of connections per unit length is suddenly reduced by a factor of $3$.
It will be  increasingly difficult for a random walker to reach large distances, and a 
decrease in the diffusion coefficient will happen at each of these characteristic lengths; thus
$\delta_n>1$ should hold for every $n$. Moreover, a comparative analysis of the structures composing Model II allows us
to obtain a better (though rather crude) lower bound. Indeed, it is possible to show that the inequality
$\delta_n>7/5$ is always true.
We can thus conclude that,  as $n$ goes to infinity, the sequences
$\delta_n, \nu_n$ and $\tau_n$ have well-defined limits ($\delta_\infty>7/5$, $\nu_\infty<0.4336$ and $\tau_\infty>12.6$, respectively),
and that an asymptotic regime emerge, in which an overall sub-diffusive behavior is modulated by log-periodic oscillations.
We then say that Model II is an {\it asymptotically diffusive self-similar} structure.

It is worth to mention that in this full self-similar structure the distribution of sites is that of the square lattice. 
Therefore, in the sense of that distribution, this substrate is a {\it trivial} fractal of dimension $d_m=2$ and without holes of 
all sizes, which could be considered the cause of sub-diffusion. 
One may ask if, regarding RW, the distribution of bonds is more important than that of sites. However, 
the related {\it  bond fractal dimension} $d_b$ is also $d_b=2$: as the counting-box linear size is increased by a 
factor of $L$ (from $L^n$ to $L^{n+1}$), 
the number of bonds increases from $N_b$ to $L^2N_b-4$ (see Fig.~\ref{mo2}, for $n=1$), which is equivalent to a $L^2$ increase factor, 
for large enough $n$. Thus, Model II exhibits logarithmic time periodicity without a (nontrivial) fractal structure.

\section{
Numerical Results
\label{nume}
}
As mentioned in Sec.~\ref{intro}, in this section, we explore diffusion using MC simulations according to models I and II.
For each model, we 
perform standard MC simulations of a single RW on the $n$-th generation basic cell. Every RW starts at the
center of symmetry of the cell. The value of $n$ is always chosen large enough to prevent the RWs from reaching the cell 
borders during the simulation. Working on this cell is thus equivalent to working with the full self-similar structure. In all
simulations the time step $\Delta t$ is set to $1$ and the hopping rate $k$ to $1/4$.

\begin{figure}[h]
\includegraphics[width=\linewidth]{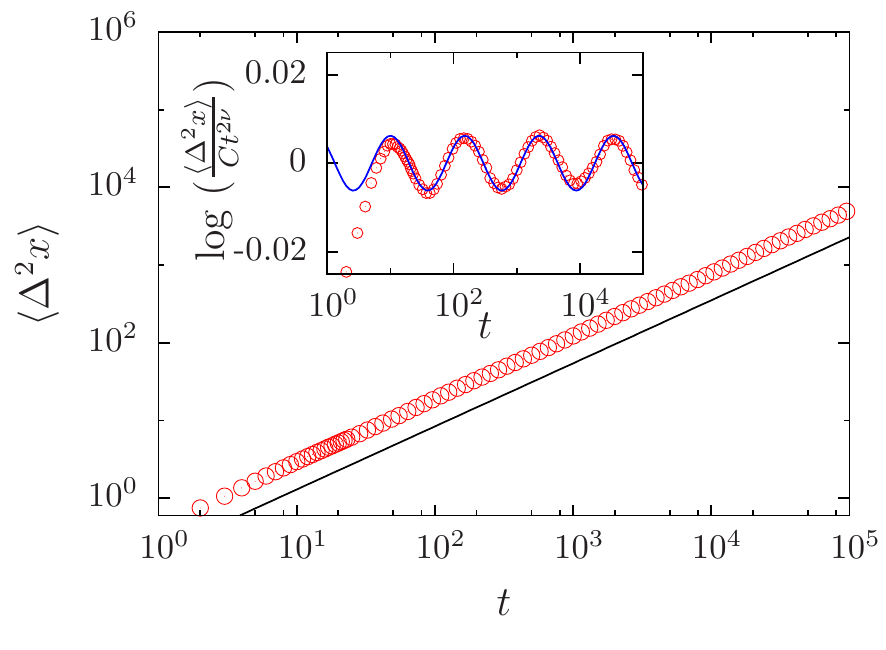}
\caption{The mean-square displacement $\displaystyle \langle \Delta^2 x \rangle$
 as a function of time for Model I. The straight line 
 has a slope $2\nu$, with $\nu\simeq 0.4057$  obtained from Eq.~(\ref{exponent}). 
The inset shows  a plot of
 $\displaystyle \log \langle \Delta^2 x \rangle / Ct^{2\nu}$ vs.
 $\log t$ for the same data, where $C$ is a properly chosen constant. The
 curvilinear line represents the first-harmonic approximation $A \sin(2\pi \log t)/\log(\tau)+\alpha)$. 
The period $\tau$ is given by Eq.(\ref{tau}). $A$ and $\alpha$ are fitted constants.
} 
\label{cruz}
\end{figure}

\begin{figure}
\includegraphics[width=\linewidth]{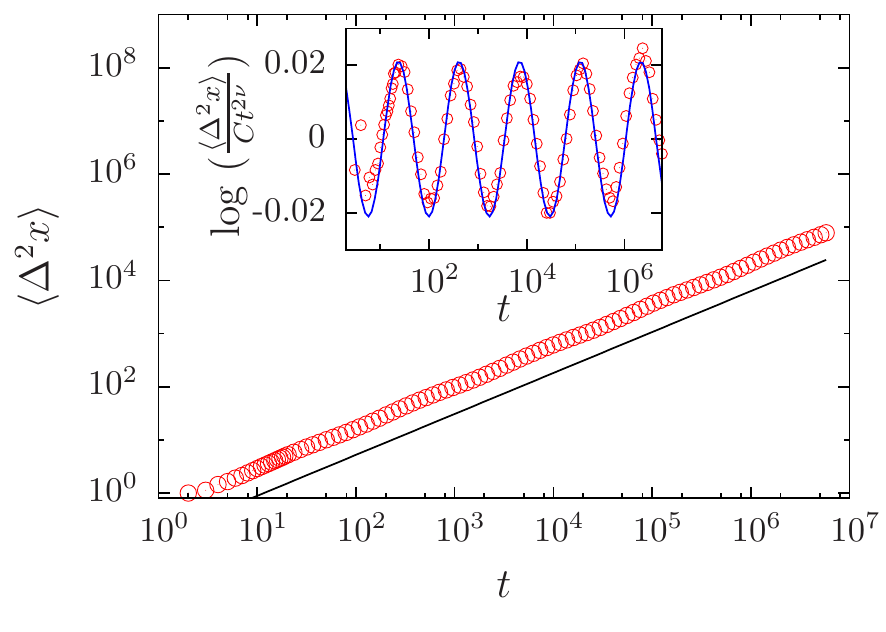}
\caption{Log-log plot of the mean-square displacement versus time for Model II. The straight line has a numerically fitted slope 
$2\nu=0.77$. The inset is a plot $\displaystyle \log \langle \Delta^2 x \rangle / Ct^{2\nu}$ 
for the same data and the curvilinear line was obtained as in Fig.~\ref{cruz}. 
The period $\tau$ is here obtained from Eq.~\ref{tau} with the numerical value of $\nu$ ($\nu_{fit}$).
}
\label{cuadra}
\end{figure}

The numerical results for the mean-square displacement are plotted as a function of time
in Figs.~\ref{cruz} and \ref{cuadra}  for Model I and Model II, respectively. It
is apparent from these figures that $\displaystyle  \langle \Delta^2 x \rangle(t)$
satisfies a modulated power law. The modulations are more easily observed in each inset,
where   
$\displaystyle\log(\langle \Delta^2x \rangle / Ct^{2\nu})$ versus
$\log(t)$, is plotted using the same data as in the corresponding main plot ($C$ is a constant chosen to have the oscillations 
centered around zero). 
In the case of Model I, the value of the RW exponent ($\nu=0.4057)$ was computed from Eq.~(\ref{exponent}), 
while for Model II it was numerically fitted ($\nu_{fit}=0.385$).
Note that, in the last case, the value is very close to the analytical approximation $\nu_{1}=0.3846$, indicating
that the convergence of the sequence is quite fast. 
The curvilinear lines are of the form 
$\displaystyle A \sin(2\pi \log(t)/\log (\tau) + \alpha )$, i.~e., the
first-harmonic approximation of a periodic function
with period $\log(\tau)$, where $A$ and
$\alpha$ are fitted parameters.  From these figures it is indeed clear that the 
predictions in Eqs.~(\ref{tau}), (\ref{exponent}), (\ref{exponent_n}) and (\ref{tau_n}) are consistent  
with the numerical findings. Note also that the amplitude of the oscillations is larger in the case of Model II, consistent
with its higher value of $\delta$.

\section{
Conclusions and discussion
\label{conclu}
}
We have studied the time evolution of a single RW on a self-similar finitely ramified structure.
The arguments employed in this work show that the time-modulated oscillatory behavior these systems exhibit
originates from finite ramification.

Schematically, at a length $L^n$, the characteristic unit cell  may
be visualized as a cage with a given number of gates. While, inside
the cage, the RW evolves normally, the situation is significantly different in the border, where
the diffusive motion is effectively slowed down due to the small number of gates.
A relevant parameter is the ratio between the number of gates and the cage perimeter.        
Because of finite ramification and self-similarity, this ratio decreases as $n$ increases, and thus,
the larger the cage, the harder it will be for the RW to find a way to get out.

The RW dynamics can thus be interpreted as a sequence of normal diffusion behaviors, each 
characterized by a diffusion coefficient, which decreases as the length-scale increases.
The oscillatory modulation amplitude of the mean-square displacement is thus
an emergent property of the transitions between two normal diffusion regimes.
Because of the substrate self-similarity, the  oscillatory behavior survive
in the long-time limit.

We have shown that there exist two wide classes of self-similar finitely ramified structures.
For pedagogical reasons, we have illustrated their properties by  studying two models: 
the snowflake fractal (Model I) and a  (trivial) fractal with a fractal dimension equal to two (Model II).
We say that Model I is perfect diffusive self-similar because, on this substrate, a single RW
follows, at all time-scales, a power law modulated by log-periodic oscillations.
In contrast, we say that Model II is asymptotically diffusive self-similar since, in this case, 
the log-periodic modulated power law regime is only reached for long enough times.
For Model I we have obtained analytically both the global RW exponent and the period of the oscillations. 
Instead, for Model II, we have shown that a convergent sequence of local  exponents and periods exists, and have calculated
its first few elements. To check the validity of our approach, MC simulations,  were also carried out for each model.  
The numerical results confirm our theoretical predictions.

We would like to stress that 
the (nontrivial) fractal character of the substrate is not a necessary condition for anomalous sub-diffusion modulated 
by log-periodic oscillations. It is a general belief that,  in a fractal, a RW becomes sub-diffusive because of 
the holes of all sizes, which hinder access to some regions. However, we have found a structure (Model II) without those
holes but leading to sub-diffusion.

Let us finally observe that, so far, we have considered  RWs that always start at the center of symmetry of the structure. 
However, as shown elsewhere \cite{lore2}, for long enough times, the diffusion became independent of the initial position,
 and the qualitative behavior of Fig.~\ref{qualita} will hold for $L^n \gg\ell$, where $\ell$ is the distance between
the initial position and that center.

In summary, for a large set of substrates, we have found that the RW sub-diffusive behavior modulated by a log-periodic
amplitude can be viewed as an emergent property of self-similarity and finite ramification. 

This work was supported by the Universidad Nacional de Mar del Plata and the 
Consejo Nacional de Investigaciones Cient\'{\i}ficas y T\'ecnicas (CONICET).


\begin{thebibliography}{99}
\bibitem{lore2}
A similar behavior can be observed in figure 7 of Ref.\cite{lore1}.
\end{thebibliography}
\end{document}